\documentstyle[preprint,prb,aps,graphics]{revtex}

\begin{document}
%\voffset=-1truein 
%\draft command makes pacs numbers print
\draft
\title{First order magnetic transition in CeFe$_2$ alloys: 
Phase-coexistence and metastability}
\author{S. B. Roy$^{1,2}$, G. K. Perkins$^1$,M. K. Chattopadhyay$^2$, 
A. K. Nigam$^3$, K. J. S. Sokhey$^2$, P.Chaddah$^2$, A. D. Caplin$^1$ and
L. F. Cohen$^1$}
\address{$^1$Blackett Laboratory, Imperial College, London SW7 2BZ, UK\\
$^2$Low Temperature Physics Laboratory,
Centre for Advanced Technology, Indore 452013, India\\
$^3$Tata Institute of Fundamental Research, Mumbai 400005, India}

\date{\today}
\maketitle
\begin{abstract}
First order ferromagnetic (FM) to antiferromagnetic (AFM) 
phase transition in doped-CeFe$_2$
alloys is studied  with micro-Hall probe technique. 
Clear  visual evidence of magnetic phase-coexistence on micrometer scales 
and the evolution 
of this phase-coexistence as a function of temperature, magnetic field  and time
across the first order FM-AFM transition is presented. 
Such phase-coexistence and metastability arise 
as natural consequence of an 
intrinsic disorder-influenced first order transition. Generality of this 
phenomena involving other classes of materials is discussed.  

\end{abstract} 

The effect of quenched disorder on a first order phase transition has been 
a subject of considerable scientific interest since late 1970's \cite{1}. 
In condensed matter physics,  
several distinct examples are known; disordered- ferroelectric 
transitions\cite{2}, precursor effects in martensitic transitions\cite{3}, 
the vortex matter phases of high temperature superconductors (HTS)\cite{4} 
and electronic phase-separation in manganites showing colossal 
magnetoresistance\cite{5}. Of these the last two areas have drawn much 
attention in recent years, although without general recognition 
that there exists some common 
underlying physics. Detailed computational studies\cite{6,7}confirm the applicability 
of early theoretical picture\cite{1} in manganites, 
and further emphasise that  phase-coexistence can  
occur in any system  in the presence of quenched disorder whenever two  
states are in competition through a first order phase transition.
Here we choose to study a simple 
binary magnetic system CeFe$_2$ and show clear  visual evidence
of magnetic phase-coexistence on micrometer scales as the system is driven 
across the entire first order antiferromagnetic (AFM) to ferromagnetic (FM)
transition. We explore how this phase-coexistence evolves as a function of 
temperature, magnetic field  and time highlighting the generic features of a 
first order FM-AFM transition. The temporal evolution of the phase-coexistence demonstrates 
that nucleation and growth can lead to percolation of a particular phase and the existence of percolative behaviour is highly dependent on  the random disorder landscape. This observation has important implications for other systems\cite{8,9} where 
percolative properties dominate macroscopic behaviour.

CeFe$_2$ is a cubic Laves phase ferromagnet (with Curie temperature
($\approx$ 230K)\cite{10}, where small substitution ($<$10 $\%$) of selected 
elements such as Co, Al, Ru, Ir, Os and Re can induce a low temperature 
AFM state with higher resistivity than the FM state\cite{11} . A giant 
magnetoresistance effect\cite{12,13} and  various memory effects\cite{14,15}
associated with the AFM-FM transition are well documented. We have chosen 
two CeFe$_2$ alloys with 4 and 5\% Ru-doping for our present work. The 
preparation and characterisation of these polycrystalline alloys have been 
described in detail in ref. 11, and samples from the same batch have been 
well-characterised\cite{11,12,15,16,17}. Neutron diffraction studies of the 
same samples revealed a discontinuous change of the unit cell volume at 
the FM-AFM transition,  confirming that it is first order\cite{17}. Bulk 
magnetic measurements were made with a vibrating sample magnetometer 
(Oxford Instruments) and a SQUID magnetometer (Quantum Design-MPMS5). 
The magnetic field profiles close to the sample surface were obtained 
using a scanning Hall probe system\cite{18} with 5 $\mu$m square InSb 
Hall sensors. The sensor was scanned at a distance of 7 $\mu$m from the 
sample surface, each image comprising of 256x256 pixels. Fields up to 
40 kOe can be applied, and the stray induction from the sensor is so 
small ($\approx$0.01 Oe) that it does not perturb the sample. 

In fig. 1, the main panel shows the global magnetisation (M) versus 
temperature (T) in an applied field (H) of 5 kOe for the 
Ce(Fe$_{0.95}$Ru$_{0.05}$)$_2$ sample (Ru-5). Two different measurement 
protocols were used: zero-field cooled (ZFC) and field-cooled cooling (FCC).
The paramagnetic(PM)-FM transition is marked by the rapid rise of M with 
decreasing T below 200K and it is thermally reversible. The FM-AFM 
transition is marked by the sharp drop in M below 90K and shows substantial
thermal hysteresis, which is an essential signature of a first order 
transition. We have obtained similar M-T curves for this sample as well as 
for the Ce(Fe$_{0.96}$Ru$_{0.04}$)$_2$   sample (Ru-4) in various H. 
Thermal hysteresis is always present in the AFM-FM transition, and broadens
with increasing H, so that when  H $>$30 kOe (15 kOe) for Ru-5 ( Ru-4 ) 
the M$_{FCC}$ (T) and M$_{ZFC}$ (T)  curves fail to merge at least down to 20K. Inset of Fig. 1 shows the schematic H-T phase diagram for the Ru-5 sample based on our M(T) measurements with T$_{NW}$(T$_{NC}$) as the temperature of the 
sharp rise (fall) in M in the ZFC (FCC) path (see Fig. 1). T$_{NC}$ is more 
closely defined as the temperature where dM/dT in the M  vs T plot changes 
sign from negative to positive. T* is the low temperature point where 
MZFC (T) and MFCC  (T) merges and T**  is the high temperature counterpart.
 A qualitatively similar phase diagram is obtained for the Ru-4 sample 
with lower characteristic temperatures.  Note that T$_{NW}$ (H) $<$T$_{NC}$ (H), i.e. the onset of nucleation of the AFM state on cooling occurs at a 
higher temperature than does nucleation of the FM phase during warming.   
This is an indication of a disorder-influenced first order transition. 
As discussed below, the sector of 
the (H,T) phase diagram shown in the inset to Fig. 1  bounded by the 
T$_{NC}$(H) ( T$_{NW}$(H) ) and T*(H) ( T**(H) ) lines  is  metastable in 
nature and susceptible to energy fluctuations. We identify 
T*(H) ( T**(H) ) as the low tempertaure (high temperature) limit of 
metastability in the free energy curves across a first order transition$^{19}$.
When the system is trapped in the metastable higher energy state it can 
be moved into the stable lower energy state by creating energy 
fluctuations such as cycling T or H- we refer to this process 
as "shattering" of the metastable state.   

We shall address now on the issue of phase-coexistence and metastability in more details by exploring the FM-AFM transition region with isothermal field variation. In such isothermal experiments one is dealing with constant energy fluctautions (coming from k$_{\beta}$T term) all along the transition region. 
In Fig.2 we present the global isothermal M-H curve of the Ru-5 sample at 60K. We also insert Hall-probe images taken at representative
fields around this M-H curve, which are highly informative. In the ascending 
H-cycle there is a sharp rise in magnetization around 19 kOe indicating the onset of AFM-FM transition. In the H-regime below 19 kOe the sample is entirely in the AFM state, and
the nature of the scanning Hall-probe field images remains same. Around 19 kOe random patches of high intensity local stray field appear in the sample, indicating the onset of FM phase in some parts of the sample. FM clusters of various size in the range of 5-20 micrometer are clearly distinguishable. The clusters grow in size and new clusters appear with further increase in H, and some of those merge to give rise to even larger clusters in the range 20-100 
micrometer and this process continues until the whole sample reaches the 
FM state. While decrteasing field from 40 kOe, along 
with bulk magnetisation the Hall-images also show distinct thermal 
hysteresis. Traces of FM clusters remain down to 15 kOe, while the sample 
was completely in the AFM state in the regime H$<$19 kOe in the ascending 
field path.  We attribute this to supercooling of the FM state across the 
first order transition. We show here that a supercooled 
state i.e. 16 kOe in the descending H cycle, is susceptible to energy 
fluctuations. A small thermal perturbation in the form of an increase in 
temperature by 10 K and then bringing the sample back to 60K again, 
markedly decreases the amount of  supercooled FM state. Note that the 
increase in T by 10 K  should drive the system towards the higher 
temperature FM phase. However, the energy fluctuation associated with 
this temperature cycling is detrimental  for the supercooled FM state. 
We have seen very same features of phase-coexistence and metastability in 
micro-Hall probe scanning images, on temperature cycling across the 
AFM-FM transition while keeping the field constant. The detail results are not shown here for the sake of conciseness. 
Supercooling  here is rationalised with the existence of a lower limit of 
metastability T*(H*) in the free energy curve  with the control parameter 
T(H)\cite{19}. We might expect to see a signature of  superheating in the 
same way. The trace of superheated AFM state would remain as a shaded 
region in an almost completely illuminated image frame.  Better resolution 
of the  images is required to reach a firm conclusion in this regard, 
and locate an upper limit of metastability temperature (field) T**(H**).

The sample used for Hall-imaging here is of dimension 2mmx1.2mmx1.2mm, and 
the images are obtained by zooming on an area of 1mm x 1mm in the central 
portion of the sample away from the edges. The whole set of experiments 
were repeated for another sample of dimension 1mmx1.2mmx1.2mm. Exactly 
the same features are observed, which  negates any dominant role of 
sample geometry. The Hall-probe images taken across the 
FM-PM transition of both the samples show a continuous 
decrease of uniformly distributed field intensity, which is  both consistent 
with a second order phase transition and confirmation that the samples 
are macroscopically chemically homogeneous. We can further rule out gross 
chemical phase separation with the results of X-ray 
diffraction\cite{11}  and  neutron scattering\cite{17} studies. 
We assert here that purely statistical quenched compositional disorder is 
at the root of the phase-coexistence observed here. Precisely this kind of 
intrinsic compositional disorder is thought to give rise to 
"tweed structure" in the vicinity of martensitic transitions\cite{3} and 
to phase-separation on sub-micrometer scale in manganites\cite{5,6,7}. 
The influence of intrinsic compositional disorder 
(through Ru-substitutions) on the critical fluctuations phenomena at the
second order PM-FM transition in the present CeFe$_2$ alloys has earlier 
been studied through detailed magnetic measurements\cite{16}.

The regions in the Ru-5 sample which go to the FM state 
first in the ascending field cycle (Fig.2) are very 
different from those which transform first to the AFM state in the descending 
field cycle (Fig.2). This is indicative of the 
local variation of the AFM-FM transition temperature T$_N$  or field H$_M$  
leading to a rough T$_N$(x,y) or  H$_M$(x,y) landscape. This distribution 
of T$_N$ and H$_M$ gives rise to the impression of global rounding of the 
transition in the bulk measurements. Our observation is in consonance 
with the disordered influenced first order transition proposed by 
Imry and Wortis\cite{1}. A very similar disorder induced rough landscape 
picture has earlier been proposed for the vortex solid melting in the HTS 
material BSCCO\cite{4}  and materials with a pre-martensitic 
transition\cite{3}.  We have already seen that the traces of the 
supercooled-FM phase remain in fields well below the 
onset temperature field  of the FM state in the 
ascending field cycle. If there was a single first order transition 
field H$_M$ ( temperature T$_N$ ) the FM clusters would have  appeared in 
the ascending field cycle in the sample first at the positions 
with relatively  low energy barrier for the nucleation of the stable FM 
phase .  Using the same argument, in the descending field cycle 
the stable AFM phase would appear first at these very points, and  these 
spots should have been the spots of lower field intensity.  However, the 
first FM-patches not only survived in the 
descending field cycle,  but some actually continued to exist  as  
supercooled metastable FM state (see Fig.2). The very same features are observed in the temperature variation measurements ( not shown here). This again emphasised a rough T$_N$(x,y)-H$_M$(x,y) landscape picture.

We provide additional evidence that the phase-coexistence regime
 bounded by T$_{NC}$(H) ( T$_{NW}$(H) )  and T*(H) ( T**(H) ) line 
(see inset of Fig.1) indeed metastable. Fig. 3 shows snap 
shots of  the temporal evolution of the  FM-phase clusters while  
undergoing  the field induced  AFM-FM transition. In these images  
T is fixed  at 60K and H is fixed at 20 kOe 
after starting from zero field in the ZFC condition. Twenty scanning 
Hall-probe images are taken over a period of 168 minutes and two 
representative images have been selected.  Significant temporal growth 
of the FM clusters  after their initial nucleation at random positions 
of  the sample is clearly visible. This is also a strong indication that 
the AFM state is superheated at least in some  regions of the sample.

We have observed qualitatively similar features of phase-coexistence and 
metastability in the Ru-4 sample both in the bulk magnetization and 
Hall-imaging studies. In this sample with less doping the 
growth of the FM phase at the onset of the AFM-FM transition, both as a 
function of T and  H, is faster than the Ru-5 sample. Growth rate can be correlated directly with the sharpness of the 
M-T and M-H curves in this sample\cite{15}. This relatively 
fast growth process along with the smaller number of nucleating clusters 
prohibited a clear-cut observation of the cluster size distributions in 
the phase-coexistence regime in the imaging experiments. 
Moreover it is now demonstrated here the way different disorder landscapes can control nucleation and growth. 
The key point is that if growth is slow enough, percolation will occur over an obnservable H (or T) interval before phase coexistence collapses. In the present study this is clearly true for the x = 0.5 sample. Percolative behaviors can be controlled by quite 
subtle changes in sample doping. The ramification to the manganite system , for example, is obvious.

In conclusion then we have imaged  FM-AFM phase-coexistence across the 
AFM-FM transition  in two Ce(Fe,Ru)$_2$ alloys. This AFM-FM transition bears 
distinct signatures of a first order phase  transition namely, 
supercooling, superheating and  time-relaxation. We have imaged the 
temporal growth of the clusters inside  the phase co-existence  regime 
for the first time and shown that this regime is quite sensitive to any 
energy fluctuations. Phase-coexistence and metastability arise as a 
consequence of  the intrinsic disorder influenced first order 
transition\cite{5,6,7}. The clusters in the present phase-coexistence 
regime have a size distribution in the range of 5-100 micrometer. 
This is larger than the  length  scale  of 0.5 $\mu$m observed previously
in manganites\cite{5,8} and discussed in the existing theories\cite{5,6,7}.
The smallest size of  clusters 
which can be detected in our present study is limited by the  resolution 
of the Hall-probe (5 $\mu$m). However, the growth
and merger of the clusters as a function of T, H and time lead naturally 
to the observation of a range of cluster size.  It will be interesting now 
to see whether such a wide scale of cluster size distribution extending to 
micrometer scales is possible within the existing class of theoretical 
models\cite{6,7}  or  whether our observations stimulate further 
theoretical refinement.   Comparing further with the rough landscape 
picture of the vortex-matter melting transition\cite{4}, our observation 
highlights the generality of the phase-coexistence phenomenon. This in 
turn points to the possibility of the key role of intrinsic disorder 
influenced first order transitions in other classes of material of 
current interest namely giant magneto-caloric materials\cite{21,22} and magnetic shape memory alloys\cite{23}. 

SBR acknowledges financial support in the form of a EPSRC visiting fellowship.

\begin{figure}
\centerline{\includegraphics{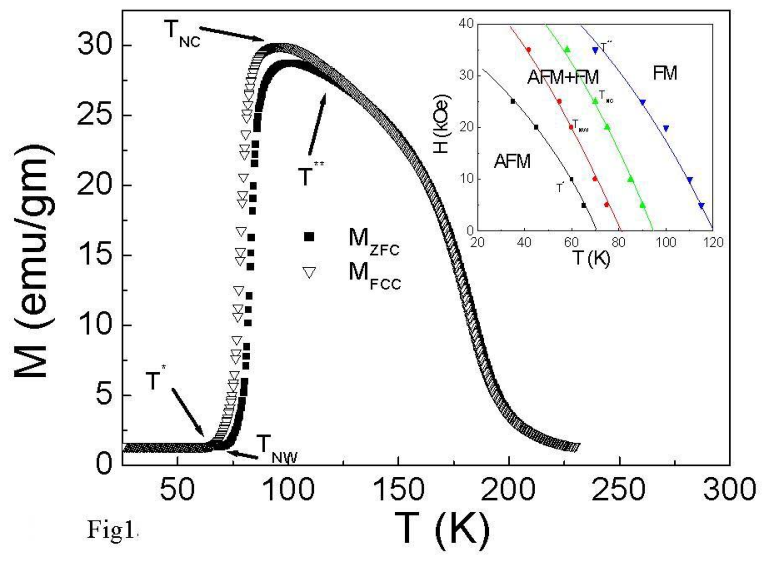}}
\caption{M versus T in an applied H of 5 kOe for the Ru-5 
sample, in ZFC and FCC mode. Inset shows the H-T phase diagram representing 
T$_{NW}$, T$_{NC}$, T* and T** (see text for their definations) as a function of H. 
The value of T* goes below 20K when H $>$ 30 kOe.}
\end{figure}

\newpage

\begin{figure}
\centerline{\includegraphics{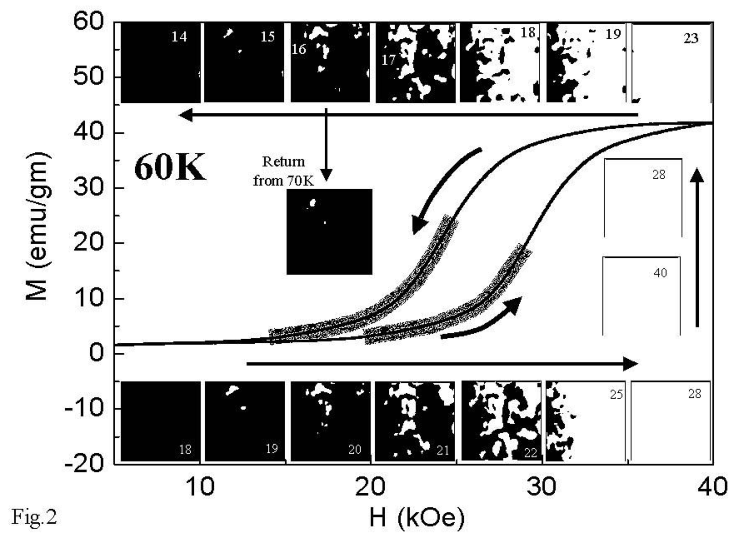}}
\caption{Isothermal M versus H plot at 60K after cooling in zero magnetic 
field.  The representative Hall probe images are inserted around 
the main figure. Starting counter-clockwise from the bottom left hand 
corner the images represent   the AFM state in the ascending field cycle 
(H=18 kOe), AFM-FM transition regime in the ascending field cycle 
( H=19, 20, 21, 22, 25, 28 kOe) , FM state ( H=40 kOe), FM-AFM transition 
regime in the descending field cycle (H=28, 23, 19, 18, 17, 16, 15 kOe) 
and  the final  AFM state (H=14 kOe) at the end of the cycle . 
Each frame covers an area of 1 x 1 mm in the central 
portion of a sample  of dimension  2 mm x 1.2 mm x 1.2mm. The 
field intensity  distribution in the images is uniform in  the 
FM state and AFM state with the intensity of the AFM state being much 
less than the FM state.  
We choose a 20\% criterion\cite{20} to highlight the onset  of the AFM-FM 
transition in the region marked by the black band. It, however,  may give the wrong 
impression that  the formation of FM state is completed by 26 kOe. 
The FM state actually goes on developing on the ascending field path until 
35 kOe, and these developments in the higher T regime  can be visualised 
 on choosing a higher ($>$20\%) threshold criterion
The frame below 16 kOe in the top row shows the effect of temperature 
cycling of 10 K on the supercooled FM state at 16 kOe. }
\end{figure}

\newpage
 
\begin{figure}
\centerline{\includegraphics{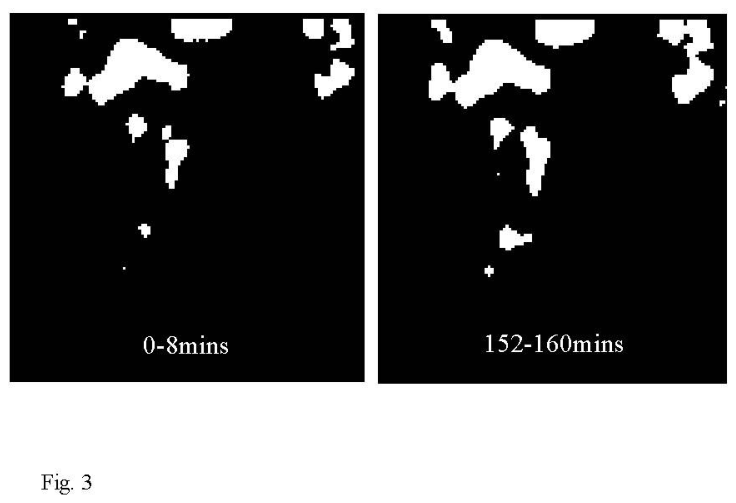}}
\caption{Images showing temporal evolution of the phase-coexistence at 
20 kOe during  the isothermal field induced AFM-FM transition at 60K.  
Two images were taken 160 minutes apart, with each image taking 8 minutes 
of experimental time to complete. The sample area scanned and the 
criterion for the colour code of the image remain same as in Fig.2. }
\end{figure}

\begin{references}
\bibitem{1}Y. Imry and M. Wortis, Phys. Rev. {\bf B19}, 3580 (1979).
\bibitem{2}H. Scjhremmer, W. Kleemann and  D. Rytz, Phys. Rev. Lett. 
{\bf 62}, 1896(1989).
\bibitem{3}S. Kartha, J. A. Krumhansl, J. P. Sethna and L. K. Wickham,
Phys. Rev. {\bf B52}, 803 (1995).  
\bibitem{4}A Soibel et al, Nature(London) {\bf 406} 283 (2000).
\bibitem{5}E. Dagotto, T. Hotta and A. Moreo, Phys. Rep. {\bf 344} 1 (2001)
; E. Dagatto, The Physics of Manganites and Related Compounds (Springer, 2003) (and references there in).
\bibitem{6}A. Moreo et al, Phys. Rev. Lett. {\bf 84}, 5568 (2000).
\bibitem{7}J. Burgy et al, Phys. Rev. Lett. {\bf 87} 277202 (2001).
\bibitem{8}M. Uehara, S. Mori, C. H. Chen and S. W. Cheong, 
Nature {\bf 399} 560 (1999).
\bibitem{9}L. Zhang et al, Science {\bf 298} 805 (2002).
\bibitem{10}L. Paolisini et al, Phys. Rev. {\bf B58} 12117 (1998) (and references therein).
\bibitem{11}S. B. Roy and B. R. Coles, J. Phys.:Condens. Matter {\bf 1} 419 (1989); Phys. Rev. {\bf B39} 9360 (1990).
\bibitem{12}H. Kunkel et al, Phys. Rev.{\bf B53} 15099 (1996).
\bibitem{13}H. Fukuda et al Phys. Rev. {\bf B63} 054405 (2001).
\bibitem{14}M. A. Manekar et al, Phys. Rev. {\bf B64} 104416 (2001).
\bibitem{15}M. K. Chattopadhyay et al. Phys. Rev.
{\bf B68} 174404(2003); K. J. S. Sokhey et al, Solid St. Commun. {\bf 129} 19 (2003).
\bibitem{16}D. Wang, H. P. Kunkel and G. Williams, 
Phys. Rev. {\bf B51} 2872 (1995).
\bibitem{17}S. J. Kennedy and B. R. Coles, J. Phys.:Condens. Matter {\bf 2} 1213 (1990).
\bibitem{18}G. K. Perkins et al Supercond. Sci. Tech. {\bf 15} 1156 (2002). 
\bibitem{19}P. M. Chaikin and T. C. Lubensky, Principles of Condensed 
Matter Physics (Cambridge University Press, Cambridge, England, 1995);P.Chaddah and S. B. Roy,
Phys. Rev. {\bf B60} 11926(1999).
\bibitem{20}Taking the intensity of the FM state as a 
reference, the AFM state is represented by  field intensity  less than 
20\% of this value (black) while white regions represent intensities 
greater than 20\% of this value. At the onset of the FM transition the 
FM-clusters of  higher field intensity appear at random position 
across the sample and they vary in size.  The whole process of this AFM-FM 
 transition both in the ascending and 
descending field cycle can be seen in a movie available at the web-site...
\bibitem{21}B. Teng et al, J.Phys.: Condens. Matter. {\bf 14} 6501(2002).
\bibitem{22}V. Pecharsky et al. Phys. Rev. Lett. {\bf 91} 197204 (2003).
\bibitem{23}W-H Wang et al. Phys. Rev. {\bf B65} 012416 (2001).  
\end{references}
\end{document}